\documentclass[11pt]{article}
\usepackage{marginnote}

\usepackage{color}
\usepackage{amsmath}
\usepackage{amsthm}
\usepackage{comment}
\usepackage{marginnote}
\allowdisplaybreaks

\usepackage{array}
\usepackage{tabularx}
\usepackage{makecell}
\usepackage{graphicx}
\usepackage{amssymb}
\usepackage[T1]{fontenc}
\usepackage[utf8]{inputenc}
\usepackage[raggedright]{titlesec}
\usepackage{blindtext}
\usepackage{commath}
\usepackage{cite}
\usepackage{caption}
\usepackage[textwidth = 7in]{geometry}
\usepackage{mathtools}
\usepackage[dvipsnames]{xcolor}
\usepackage{enumitem}
\usepackage{amsmath}
\usepackage{geometry}
\usepackage{lipsum}  
\usepackage{thmtools}
\usepackage{setspace}
\titleformat{\paragraph}[hang]{\normalfont\normalsize\bfseries}{\theparagraph}{1em}{}
\titlespacing*{\paragraph}{0pt}{3.25ex plus 1ex minus .2ex}{0.5em}
\makeatletter
\DeclareRobustCommand{\change}{%
	\@bsphack
	\leavevmode
	\color{magenta}%
	\@esphack
}
\DeclareRobustCommand{\stopchange}{%
	\@bsphack
	\normalcolor
	\@esphack
}
\makeatother

\stepcounter{tocdepth}

\usepackage{hyperref}
\hypersetup{
  colorlinks   = true, 	
  urlcolor     = blue, 	
  linkcolor    = blue, 	
  citecolor   = magenta 	
}

\title{
	\vskip-1.3cm
One-dimensional Coulomb Problem in GUP Formalism \\
}
\author{
         S. Zarrinkamar$^{1}$\footnote {saber.zarrinkamar@iua.ac.ir, zarrinkamar.s@gmail.com, ORCID: \href{http://orcid.org/0000-0001-9128-4624}{0000-0001-9128-4624}},
		\\  [1ex]
	\small
	$^1$\,Departament of Basic Sciences, Ga. C., Islamic Azad University, Garmsar, Iran\\  [1ex]
}

\begin{document}
	
	\maketitle

\begin{abstract}
We investigate the one-dimensional Coulomb problem on the positive half-line for a fourth-order Schr\"{o}dinger equation generated by a commonly used realization of the Generalized Uncertainty Principle (GUP). The problem is treated directly in position space by a higher-order Bethe--Ansatz construction, with the wave function represented as a polynomial multiplied by an exponential factor. The resulting residue conditions yield an analytic quantization condition and explicit polynomial solutions for the first three bound states. We identify the branch that is continuously connected to the ordinary Coulomb problem and show that its energies, decay constants, polynomial factors, and Bethe--Ansatz roots recover the ordinary half-line Coulomb results in the vanishing-deformation limit. On this Coulomb-connected branch, the deformation produces a lower admissibility bound on the principal quantum number, while arbitrarily high quantum numbers remain admissible. We also discuss the physical interpretation of the deformation strength: for ordinary microscopic systems, the weak-GUP regime is the conservative expectation in Planck-scale motivated models, whereas intermediate and strong regimes are primarily theoretical regimes in the present analysis. Since the differential equation is truncated at first order in the GUP parameter, quantitative predictions outside the weak-deformation regime should be interpreted with care.
\end{abstract}


	\section{Introduction}
The existence of a Minimal Length (ML) at Planck scale is predicted by many theories such as quantum gravity \cite {Maggiore, Hossenfelder, Ali}, black hole physics \cite {Scardigli, Vagenas PRD 2004}, double special relativity \cite{Cortes} as well as string theory \cite{Konishi} and can be considered as the common interface of all of them.  
Such a ML corresponds to a Generalized Uncertainty Principle (GUP) or equivalently generalized wave equations \cite {Kempf, Vagenas 2009, Review 2015}. The GUP problem has already been studied in connection with gravity and quantum gravity \cite {Vagenas PRL 2008, Casadio 2015, Singleton, Lambiase}, black hole physics \cite {Bargueno}and experimental data \cite {Vagenas PLB 2021, Matt 2023, Vagenas 2024}. The problem has been very recently studied in connection with modern and multidisciplinary subjects as well \cite {Matt 2022, Jusufi 2023}. The rather complicated structure of the problem in such cases is perhaps the main and for sure the first challenge \cite {Review 2015, Vagenas 2010 Relativistic, Nozari Pedram, Bishop 2023}. To analyze the problem, various exact or approximate approaches have been proposed for both nonrelativistic and relativistic cases, which consider only simple cases \cite {Khorram, Bosso 2021 Harmonic, PLB 2024}. \\
The problem in dealing with the GUP modified wave equations is that the equations become more complicated and that such forms are not investigated sufficiently. To be more precise, for example, the schr\"{o}dinger equation for a particular case of GUP, which is physically well motivated and frequently considered in the literature, appears in the form of a fourth-order differential equation which has been only solved for very simple cases, e.g. constant terms including the potential well, step, etc \cite {Vagenas 2010 Relativistic, Nozari Pedram, Khorram}. This is rather obvious, since our common techniques, such as factorization, integral transforms, supersymmetry quantum mechanics and others, which work quite well for the ordinary schr\"{o}dinger equation, have been rarely generalized for or checked with the higher-order differential equations with variable coefficients. Therefore, we have to either use new methods, or generalize our old ideas to the GUP problem to understand the structure better and deeper \cite{Cooper,Junker}. \\
On the other hand, the one-dimensional Coulomb problem requires care because the singularity at the origin leads to nontrivial boundary-condition and self-adjointness issues. The ordinary one-dimensional Coulomb problem is a well-established problem in the literature, beginning with the work of Loudon and followed by several analyses of its spectrum, singularity, and admissible boundary conditions \cite{1959,1966,1969 roberts,1980 gomes,1987,1996 Kurasov,1997,1999,2016 Loudon,2019}. In particular, the singularity at the origin has been shown to require careful treatment of the left and right sectors and of the corresponding boundary conditions \cite{2011,2019 Calcada}. For the positive-half-line realization studied here, with $z>0$ and the regular boundary condition $y(0)=0$, the ordinary Coulomb equation is exactly solvable and is equivalent to the familiar $\ell=0$ radial Coulomb problem. This half-line realization and its spectral properties have been discussed explicitly in recent mathematical analyses \cite{Fassari2025,Gadella2025,Fassari2026}. We therefore emphasize that the undeformed Coulomb problem and its exact spectrum are established results and are used here as the benchmark for the GUP deformation, rather than being claimed as a new solution. Related one-dimensional Coulomb-like problems in deformed spaces with minimal length have also been studied previously \cite{2006 ML,2016 ML}. \\
Here, we  will use the ingenious idea of the Bethe-ansatz approach for our case of fourth-order GUP-modified one-dimensional schr\"{o}dinger equation \cite {Zhang 2012 JPA, Zhang 2013, Zhang 2017}. In this way, which will be presented quite summarized, we first review a form of GUP which is well motivated and frequently used in the literature. We next review the novel formulation of Zhang, originally proposed in quantum optics, and finally apply the method to our case and thereby report the solutions. We give our concluding remarks in the last section and comment on the present status of the problem in the existing literature and hopefully, propose some ideas to further studies in the field, which, to the best of our knowledge, have not been investigated yet.

\section{A GUP Formalism}
We consider the commutation relation \cite {Maggiore, Review 2015, Nozari Pedram} 
\begin{equation}
[x,p]=i\hbar(1+\beta p^2),
\end{equation}
which corresponds to the GUP
\begin{equation}
\Delta x \Delta p \geq\frac{\hbar}{2} \left (1+ \beta(\Delta p)^2+\gamma \right ),
\end{equation}
where $x$ and $p$ denote the position and momentum operators, $m>0$ is the particle mass,
$\hbar$ is the reduced Planck constant, $\beta\geq0$ is the GUP deformation parameter,
and $\gamma=\beta\langle p\rangle^2$ for the state under consideration. It should be, however, stressed that different operators might yield the same commutation relation \cite {Bishop 2023}, meaning that only commutators are not enough to determine the physics of the system. Let us now choose the generalized operator form
\begin{equation}
x\rightarrow x,  \ \ \ and \ \ \ p \rightarrow p\left (1+\frac{1}{3}\beta p^2 \right). 
\end{equation}
Up to first order in $\beta $, the modified equation due to GUP appears as

\begin{equation}
\frac {\beta\hbar^4}{3m}\frac {d^4y(x)}{dx^4}-\frac{\hbar^2}{2m}\frac {d^2y(x)}{dx^2}+V(x)y(x)=Ey(x). 
\label{GUPeq}
\end{equation}

For the sake of simplicity and without loss of generality, 

\section{The Bethe-Ansatz Approach for a Fourth-Order Differential Equation}
Let us consider the fourth-order differential equation in the form
\begin{equation}
H\psi(z)=0,
\end{equation}
where
\begin{equation}
H=P_5(z)\frac{d^4}{dz^4}+P_4(z)\frac{d^3}{dz^3}
+P_3(z)\frac{d^2}{dz^2}+P_2(z)\frac{d}{dz}+P_1(z),
\end{equation}
and the polynomial coefficients can in general have the form
\begin{equation}
P_5(z)=\sum_{k=0}^{5}a_kz^k,\quad
P_4(z)=\sum_{k=0}^{4}b_kz^k,\quad
P_3(z)=\sum_{k=0}^{3}c_kz^k,\quad
P_2(z)=\sum_{k=0}^{2}d_kz^k,\quad
P_1(z)=\sum_{k=0}^{1}e_kz^k.
\end{equation}
Proposing the polynomial solution
	\begin{equation}  \label{ansatz}
\psi(z)=u(z)=\prod_{i=1}^n(z-z_i), \qquad n=1,2,\cdots ,
	\end{equation}
after a lengthy algebra, the logarithmic derivatives of $u$ can be written in terms of the roots as
	\begin{equation}
		\begin{gathered}
\frac{u'}{u}=\sum_{i=1}^n\frac{1}{z-z_i},\\
\frac{u''}{u}=\sum_{i=1}^n\frac{1}{z-z_i}
\sum_{\substack{j=1\\j\neq i}}^n\frac{2}{z_i-z_j},\\
\frac{u^{(3)}}{u}=\sum_{i=1}^n\frac{1}{z-z_i}
\sum_{\substack{l,j=1\\l\neq j,\ l,j\neq i}}^n
\frac{3}{(z_i-z_l)(z_i-z_j)},\\
\frac{u^{(4)}}{u}=\sum_{i=1}^n\frac{1}{z-z_i}
\sum_{\substack{p,l,j=1\\p,l,j\neq i\\p\neq l,\ p\neq j,\ l\neq j}}^n
\frac{4}{(z_i-z_p)(z_i-z_l)(z_i-z_j)}.
		\end{gathered}
	\end{equation}
The residue conditions obtained by requiring the apparent poles at $z=z_i$ to vanish provide the Bethe-Ansatz equations. In the present fourth-order problem, it is therefore not necessary to use the second-order polynomial differential equation; the fourth-order structure must be retained throughout the calculation.

\section{Solution to the GUP-Modified Coulomb (Gravitational) Potential}
We consider the GUP modified Schr\"{o}dinger equation with an interaction of the form 
	\begin{equation} 
V(z)=-\frac{b}{z}, \ \ \ \ b, z>0,
	\end{equation}
where the problem is defined on the positive half-line and the regular Coulomb boundary condition is imposed at the origin. For a bound state we write
\begin{equation}
E=-\epsilon, \qquad \epsilon>0.
\end{equation}
The equation then reads
	\begin{equation}
\frac{\beta\hbar^4}{3m}y^{(4)}(z)-\frac{\hbar^2}{2m}y''(z)
-\frac{b}{z}y(z)=-\epsilon y(z).
	\end{equation}
For later convenience, define
\begin{equation}\label{Sdefinition}
S=\sqrt{3(3-16\beta m\epsilon)}.
\end{equation}
The exponential factor has two algebraic branches. The branch which has a smooth ordinary-Coulomb limit as $\beta\rightarrow0$ is
\begin{equation}
a^2=\frac{3-S}{4\beta\hbar^2},
\qquad
a=\frac{\sqrt{3-S}}{2\sqrt{\beta}\hbar}.
\label{abranch}
\end{equation}
Indeed, on this branch $S\rightarrow3$ as $\beta\rightarrow0$, and therefore $a$ remains finite. The other sign,
\begin{equation}
a^2=\frac{3+S}{4\beta\hbar^2},
\end{equation}
has $a^2\sim3/(2\beta\hbar^2)$ for $\beta\rightarrow0$ and does not reproduce the ordinary Coulomb spectrum. It is consequently not included in the physical Coulomb branch considered below.

Applying the transformation
\begin{equation}
         	\begin{gathered}
y(z)=u(z)e^{-az}
=\prod_{i=1}^n(z-z_i)e^{-az},
         	\end{gathered}
\end{equation}
and multiplying the resulting equation by $24m\beta^2$, one obtains
	\begin{equation}\label{transformed}
		\begin{gathered}
-24\beta^2mb\,u(z)
+4\beta^{3/2}\hbar S\sqrt{3-S}\,z\,u'(z)\\
+12\beta^2\hbar^2(2-S)z\,u''(z)
-16\beta^{5/2}\hbar^3\sqrt{3-S}\,z\,u^{(3)}(z)\\
+8\beta^3\hbar^4z\,u^{(4)}(z)=0.
		\end{gathered}
	\end{equation}
The choice of $a$ above has removed the term proportional to $z\,u(z)$.

Proposing the solution \eqref{ansatz} and dividing both sides by it, we obtain
	\begin{equation} 
		\begin{gathered}
24\beta^2mb=
\left(8\beta^3\hbar^4z\right)
\sum_{i=1}^n\frac{1}{z-z_i}
\sum_{\substack{p,l,j=1\\p,l,j\neq i\\p\neq l,\ p\neq j,\ l\neq j}}^n
\frac{4}{(z_i-z_p)(z_i-z_l)(z_i-z_j)}\\
-\left(16\beta^{5/2}\hbar^3\sqrt{3-S}\,z\right)
\sum_{i=1}^n\frac{1}{z-z_i}
\sum_{\substack{l,j=1\\l,j\neq i,\ l\neq j}}^n
\frac{3}{(z_i-z_l)(z_i-z_j)}\\
+\left(12\beta^2\hbar^2(2-S)z\right)
\sum_{i=1}^n\frac{1}{z-z_i}
\sum_{\substack{j=1\\j\neq i}}^n
\frac{2}{z_i-z_j}\\
+\left(4\beta^{3/2}\hbar S\sqrt{3-S}\,z\right)
\sum_{i=1}^n\frac{1}{z-z_i}.
		\end{gathered}
	\end{equation}
Therefore, the residue at $z=z_i$ is obtained as
	\begin{equation} \label {Residues}
		\begin{gathered}
8\beta^3\hbar^4z_i
\sum_{\substack{p,l,j=1\\p,l,j\neq i\\p\neq l,\ p\neq j,\ l\neq j}}^n
\frac{4}{(z_i-z_p)(z_i-z_l)(z_i-z_j)}\\
-16\beta^{5/2}\hbar^3z_i\sqrt{3-S}
\sum_{\substack{l,j=1\\l,j\neq i,\ l\neq j}}^n
\frac{3}{(z_i-z_l)(z_i-z_j)}\\
+12\beta^2\hbar^2(2-S)z_i
\sum_{\substack{j=1\\j\neq i}}^n
\frac{2}{z_i-z_j}
+4\beta^{3/2}\hbar S\sqrt{3-S}\,z_i=0.
		\end{gathered}
	\end{equation}
For distinct roots, all the indices appearing in the multiple sums are understood to be pairwise distinct. Repeated roots can occur at limiting parameter values and should then be treated directly in the polynomial equation rather than by the simple-pole residue formula.

The energy relation follows from the constant term at large $z$. Since
\begin{equation}
\frac{u'}{u}=\frac{n}{z}+O(z^{-2}),
\end{equation}
whereas the terms containing $u''$, $u^{(3)}$ and $u^{(4)}$ decrease at least as $z^{-1}$ after multiplication by their respective coefficients proportional to $z$, the constant part of the equation gives
	\begin{equation} \label {energy}
-24\beta^2mb+
4n\beta^{3/2}\hbar S\sqrt{3-S}=0.
	\end{equation}
Thus,
\begin{equation} \label {energy 2}
n\hbar S\sqrt{3-S}=6mb\sqrt{\beta}.
\end{equation}
This is the energy quantization condition on the physical Coulomb branch.

Using
\begin{equation}
S^2=3(3-16\beta m\epsilon),
\end{equation}
the binding energy can be written as
\begin{equation} \label {epsilonS}
\epsilon_n=\frac{9-S_n^2}{48\beta m},
\end{equation}
where $S_n$ is determined by
\begin{equation} \label {Scondition}
S_n^2(3-S_n)
=\frac{36\beta m^2b^2}{n^2\hbar^2}.
\end{equation}
The physical Coulomb branch is the solution
\begin{equation}
2\leq S_n<3,
\end{equation}
with $S_n\rightarrow3$ when $\beta\rightarrow0$. The second real solution, when it exists, lies in $0<S_n\leq2$ and belongs to the non-Coulomb branch because it gives
$a^2\sim O(1/\beta)$ and $\epsilon_n\sim3/(16\beta m)$ as $\beta\rightarrow0$. It is therefore not counted as an ordinary Coulomb bound state.

Equation \eqref {Scondition} also gives a direct restriction on the quantum number. On the physical interval $2\leq S_n<3$, the function
\begin{equation}
f(S)=S^2(3-S)
\end{equation}
decreases monotonically from $f(2)=4$ to $f(3)=0$. Hence a physical Coulomb-branch solution exists only if
\begin{equation} \label {restriction}
\frac{36\beta m^2b^2}{n^2\hbar^2}\leq4,
\end{equation}
or equivalently
\begin{equation} \label {restriction2}
\beta\leq\frac{n^2\hbar^2}{9m^2b^2}.
\end{equation}
Therefore, for a fixed positive $\beta$, the admissible principal quantum numbers satisfy
\begin{equation} \label {nmin}
n\geq n_{\min}
=\max\left\{1,\left\lceil\frac{3mb\sqrt{\beta}}{\hbar}\right\rceil\right\}.
\end{equation}
Thus the GUP deformation produces a lower cutoff in $n$: sufficiently low-lying Coulomb states cease to belong to the physical Coulomb branch when $\beta$ is sufficiently large. In the ordinary limit $\beta\rightarrow0$, the restriction disappears and all
$n=1,2,\ldots$ are recovered.

At the limiting value
\begin{equation}
\beta=\frac{n^2\hbar^2}{9m^2b^2},
\end{equation}
one has $S_n=2$. In this case the standard distinct-root residue representation may become degenerate; the corresponding polynomial solution should be understood as the limiting solution of the differential equation. For $\beta<n^2\hbar^2/(9m^2b^2)$, one has $2<S_n<3$ on the physical branch.

It is useful to introduce the dimensionless GUP coupling
\begin{equation}
\alpha_G=\frac{\beta m^2b^2}{\hbar^2}.
\end{equation}
Then the condition for the physical Coulomb branch becomes simply
\begin{equation}
\alpha_G\leq\frac{n^2}{9},
\end{equation}
and
\begin{equation}
n_{\min}=\max\{1,\lceil3\sqrt{\alpha_G}\rceil\}.
\end{equation}
For example, if $0<\alpha_G\leq1/9$, the ground state $n=1$ is allowed; if
$1/9<\alpha_G\leq4/9$, the lowest admissible state is $n=2$; if
$4/9<\alpha_G\leq1$, it is $n=3$, and so on.

To summarize, the procedure is the following. For a given integer $n$, one first solves \eqref {Scondition} on the physical interval $2\leq S_n<3$, then obtains the energy from \eqref {epsilonS}. The energy is subsequently inserted into the residue relations \eqref {Residues}, which determine the roots $z_i$. Once the roots are obtained, the wave function follows from
\eqref {ansatz}. The condition \eqref {restriction2} is therefore an admissibility condition for the Coulomb-branch bound state, not an upper bound on the number of states.

As a consistency check, expanding the physical solution for small $\beta$ gives
\begin{equation}
\epsilon_n=
\frac{mb^2}{2\hbar^2n^2}
+\frac{\beta m^3b^4}{\hbar^4n^4}
+O(\beta^2),
\end{equation}

\section{First Three Bound States: Detailed Calculations}

We now work out the first three physical states explicitly. These calculations serve as a direct check of the general residue equation~\eqref{Residues} and make the Coulomb limit transparent.

\subsection{The $n=1$ bound state}

For $n=1$, regularity at the origin requires the single root to be
\begin{equation}
z_1=0.
\end{equation}
Hence
\begin{equation}
u_1(z)=z,
\qquad u_1'=1,
\qquad u_1''=u_1'''=u_1''''=0.
\end{equation}
Substitution into the transformed fourth-order equation gives
\begin{equation}
-24\beta^2mbz+4\beta^{3/2}\hbar S_1\sqrt{3-S_1}\,z=0.
\end{equation}
Since this must hold for every $z>0$,
\begin{equation}
\hbar S_1\sqrt{3-S_1}=6mb\sqrt{\beta},
\end{equation}
which is exactly the general quantization condition for $n=1$. The wave function is therefore
\begin{equation}
y_1(z)=\mathcal N_1 z e^{-a_1z},
\qquad
a_1^2=\frac{3-S_1}{4\beta\hbar^2},
\end{equation}
with energy
\begin{equation}
E_1=-\frac{9-S_1^2}{48\beta m}.
\end{equation}
The Coulomb-connected branch is $2\le S_1<3$, and consequently
\begin{equation}
\beta\leq\frac{\hbar^2}{9m^2b^2}.
\end{equation}

\subsection{The $n=2$ bound state}

For $n=2$, write
\begin{equation}
z_1=0,
\qquad
u_2(z)=z(z-z_2)=z^2-z_2z.
\end{equation}
Its derivatives are
\begin{equation}
u_2'=2z-z_2,
\qquad u_2''=2,
\qquad u_2'''=u_2''''=0.
\end{equation}
Substitution into the transformed equation gives
\begin{align}
0={}&-24\beta^2mb(z^2-z_2z)
+4\beta^{3/2}\hbar S_2\sqrt{3-S_2}\,z(2z-z_2)\notag\\
&+24\beta^2\hbar^2(2-S_2)z.
\end{align}
The coefficient of $z^2$ is
\begin{equation}
-24\beta^2mb+8\beta^{3/2}\hbar S_2\sqrt{3-S_2}=0,
\end{equation}
so that
\begin{equation}
2\hbar S_2\sqrt{3-S_2}=6mb\sqrt{\beta},
\end{equation}
which is the $n=2$ quantization condition. The coefficient of $z$ gives
\begin{equation}
24\beta^2mbz_2
-4\beta^{3/2}\hbar S_2\sqrt{3-S_2}\,z_2
+24\beta^2\hbar^2(2-S_2)=0.
\end{equation}
Using the quantization condition to eliminate $mb$ yields
\begin{equation}
z_2=\frac{6\hbar\sqrt{\beta}(S_2-2)}{S_2\sqrt{3-S_2}}.
\end{equation}
Thus
\begin{equation}
y_2(z)=\mathcal N_2z(z-z_2)e^{-a_2z},
\qquad
a_2^2=\frac{3-S_2}{4\beta\hbar^2},
\end{equation}
and
\begin{equation}
E_2=-\frac{9-S_2^2}{48\beta m}.
\end{equation}
The Coulomb-branch existence condition is
\begin{equation}
\beta\leq\frac{4\hbar^2}{9m^2b^2}.
\end{equation}
At the endpoint $S_2=2$, one obtains $z_2=0$, so the two roots coalesce at the origin and the limiting polynomial is $u_2=z^2$. The limiting polynomial is regular; only the generic simple-pole residue representation becomes inapplicable exactly at the coalescence.

\subsection{The $n=3$ bound state}

For the third state, take the monic cubic polynomial
\begin{equation}
u_3(z)=z^3+c_2z^2+c_1z,
\end{equation}
where the constant term vanishes because the wave function obeys the regular half-line condition $y(0)=0$. Its derivatives are
\begin{equation}
u_3'=3z^2+2c_2z+c_1,
\qquad
u_3''=6z+2c_2,
\qquad
u_3'''=6,
\qquad
u_3''''=0.
\end{equation}
Substitution into the transformed equation gives a polynomial whose coefficients of $z^3,z^2,z$ must vanish separately:
\begin{align}
[z^3]:
&\quad -24\beta^2mb+12\beta^{3/2}\hbar S_3\sqrt{3-S_3}=0,
\label{n3z3}
\end{align}
\begin{align}
[z^2]:
&\quad 8\beta^{3/2}\hbar S_3\sqrt{3-S_3}\,c_2
+72\beta^2\hbar^2(2-S_3)
-24\beta^2mb\,c_2=0,
\label{n3z2}
\end{align}
\begin{align}
[z]:
&\quad 4\beta^{3/2}\hbar S_3\sqrt{3-S_3}\,c_1
+-24\beta^2\hbar^2(S_3-2)c_2\notag\\
&\qquad -24\beta^2mb\,c_1
-96\beta^{5/2}\hbar^3\sqrt{3-S_3}=0.
\label{n3z1}
\end{align}
Equation~\eqref{n3z3} gives
\begin{equation}
3\hbar S_3\sqrt{3-S_3}=6mb\sqrt{\beta},
\end{equation}
which is precisely the general quantization condition for $n=3$. Using it in Eq.~\eqref{n3z2} gives
\begin{equation}
c_2=-\frac{18\hbar\sqrt{\beta}(S_3-2)}{S_3\sqrt{3-S_3}}.
\end{equation}
Substitution of this result and the same quantization condition into Eq.~\eqref{n3z1} then gives
\begin{equation}
c_1=-\frac{6\beta\hbar^2(11S_3^2-42S_3+36)}{S_3^2(S_3-3)}.
\end{equation}
Therefore the cubic can be written in factored form as
\begin{equation}
u_3(z)=z(z-z_2)(z-z_3),
\end{equation}
where
\begin{equation}
z_{2,3}=\frac{\hbar\sqrt{\beta}}{S_3\sqrt{3-S_3}}
\left[9(S_3-2)\mp\sqrt{3(5S_3^2-24S_3+36)}\right].
\label{bethe_roots_check}
\end{equation}
For $2<S_3<3$, the two nonzero roots are distinct. However, they are not both positive throughout the whole Coulomb-connected interval. The smaller root is negative for

\begin{equation}
2<S_3<\frac{21+3\sqrt{5}}{11}\approx2.51893,
\end{equation}

and crosses the origin at $S_3=(21+3\sqrt{5})/11$. Both nonzero roots are positive for
$ (21+3\sqrt{5})/11<S_3<3$. This does not invalidate the polynomial solution on the half-line; it means that, in part of the parameter range, one of the polynomial zeros lies outside the physical domain $z>0$. The third-state wave function is consequently
\begin{equation}
y_3(z)=\mathcal N_3 z(z-z_2)(z-z_3)e^{-a_3z},
\qquad
a_3^2=\frac{3-S_3}{4\beta\hbar^2},
\end{equation}
with
\begin{equation}
E_3=-\frac{9-S_3^2}{48\beta m},
\qquad
\beta\leq\frac{\hbar^2}{m^2b^2}.
\end{equation}
The last inequality follows from the general condition $\beta\le n^2\hbar^2/(9m^2b^2)$ with $n=3$. The direct coefficient matching above provides an independent verification of the residue construction for the first genuinely cubic polynomial.

\section{Special-Case Checks and the Vanishing-$\beta$ Limit}

The polynomial solutions provide direct checks of the transformed fourth-order equation~\eqref{transformed} and of the Bethe--Ansatz quantization condition~\eqref{app:A59}.  We
first examine the constant, linear, and quadratic cases.  We then give an
independent derivation of the ordinary one-dimensional Coulomb problem by
setting $\beta=0$ before solving the differential equation.  This last
calculation provides a stringent check on the GUP spectrum and on the
polynomial structure of the first three states. The ordinary half-line Coulomb
problem used in this check is an established exactly solvable problem in the
literature \cite{1959,2011,2019 Calcada,Fassari2025,Gadella2025,Fassari2026}; our purpose here is to verify explicitly
that the present GUP construction reduces to that known result as
$\beta\rightarrow0$.

For $n=0$ let
\[\refstepcounter{equation}\tag{\theequation}
u_0(z)=C_0.
\]
All derivatives vanish, so the transformed equation~\eqref{transformed} reduces immediately to
\[\refstepcounter{equation}\tag{\theequation}
-24\beta^2mb\,C_0=0.
\]
For $\beta>0$, $m>0$, and $b>0$, the only solution is
\[\refstepcounter{equation}\tag{\theequation}
C_0=0.
\]
Thus no nontrivial constant polynomial occurs.  This is also required by
the regular half-line boundary condition $y(0)=0$: a nonzero constant
would give $y(0)\neq0$.  The physical polynomial sequence therefore begins
with $n=1$.

Take
\[\refstepcounter{equation}\tag{\theequation}
u_1(z)=z,
\qquad
u_1'=1,
\qquad
u_1''=u_1'''=u_1''''=0.
\]
The transformed equation~\eqref{transformed} becomes
\[\refstepcounter{equation}\tag{\theequation}
-24\beta^2mb\,z
+4\beta^{3/2}\hbar S\sqrt{3-S}\,z=0.
\]
For a nontrivial state,
\[\refstepcounter{equation}\tag{\theequation}
\hbar S_1\sqrt{3-S_1}
=
6mb\sqrt{\beta}.
\]
Squaring,
\[\refstepcounter{equation}\tag{\theequation}
S_1^2(3-S_1)
=
\frac{36\beta m^2b^2}{\hbar^2}.
\]
Since
\[\refstepcounter{equation}\tag{\theequation}
S_1^2=9-48\beta m\epsilon_1,
\]
the energy is
\[\refstepcounter{equation}\tag{\theequation}
E_1=-\epsilon_1
=
-\frac{9-S_1^2}{48\beta m}.
\]
On the Coulomb-connected branch $2\leq S_1<3$, the endpoint $S_1=2$
gives
\[\refstepcounter{equation}\tag{\theequation}
4=\frac{36\beta m^2b^2}{\hbar^2},
\]
hence
\[\refstepcounter{equation}\tag{\theequation}
\beta\leq\frac{\hbar^2}{9m^2b^2}.
\]
The wavefunction uses the Coulomb-connected exponential branch~\eqref{abranch}:
\[\refstepcounter{equation}\tag{\theequation}
y_1(z)=\mathcal N_1ze^{-a_1z},
\qquad
a_1^2=\frac{3-S_1}{4\beta\hbar^2}.
\]
For $\beta\to0$ write $S_1=3-\delta_1$.  Then
\[\refstepcounter{equation}\tag{\theequation}
(3-\delta_1)^2\delta_1
=
\frac{36\beta m^2b^2}{\hbar^2},
\]
so
\[\refstepcounter{equation}\tag{\theequation}
\delta_1
=
\frac{4\beta m^2b^2}{\hbar^2}
+O(\beta^2).
\]
Consequently,
\[\refstepcounter{equation}\tag{\theequation}
a_1^2
=
\frac{m^2b^2}{\hbar^4}+O(\beta),
\qquad
a_1\to\frac{mb}{\hbar^2},
\]
and
\[\refstepcounter{equation}\tag{\theequation}
E_1\to-\frac{mb^2}{2\hbar^2}.
\]

Take
\[\refstepcounter{equation}\tag{\theequation}
u_2(z)=z(z-z_2)=z^2-z_2z,
\]
so
\[\refstepcounter{equation}\tag{\theequation}
u_2'=2z-z_2,
\qquad
u_2''=2,
\qquad
u_2'''=u_2''''=0.
\]
Substitution gives
\[\refstepcounter{equation}\tag{\theequation}
-24\beta^2mb(z^2-z_2z)
+4\beta^{3/2}\hbar S\sqrt{3-S}\,z(2z-z_2)
+24\beta^2\hbar^2(2-S)z=0.
\]
The $z^2$ coefficient yields the $n=2$ specialization of~\eqref{app:A59}:
\[\refstepcounter{equation}\tag{\theequation}
-24\beta^2mb
+8\beta^{3/2}\hbar S\sqrt{3-S}=0,
\]
or
\[\refstepcounter{equation}\tag{\theequation}
\hbar S_2\sqrt{3-S_2}=3mb\sqrt{\beta}.
\]
Equivalently,
\[\refstepcounter{equation}\tag{\theequation}
2\hbar S_2\sqrt{3-S_2}=6mb\sqrt{\beta},
\]
and
\[\refstepcounter{equation}\tag{\theequation}
S_2^2(3-S_2)
=
\frac{9\beta m^2b^2}{\hbar^2}.
\]
The $z$ coefficient is
\[\refstepcounter{equation}\tag{\theequation}
24\beta^2mb\,z_2
-4\beta^{3/2}\hbar S_2\sqrt{3-S_2}\,z_2
+24\beta^2\hbar^2(2-S_2)=0.
\]
Using the quantization condition gives
\[\refstepcounter{equation}\tag{\theequation}
12\beta^2mb\,z_2
+24\beta^2\hbar^2(2-S_2)=0,
\]
and hence
\[\refstepcounter{equation}\tag{\theequation}
z_2
=
\frac{2\hbar^2(S_2-2)}{mb}.
\]
Eliminating $mb$ gives
\[\refstepcounter{equation}\tag{\theequation}
z_2=
\frac{6\hbar\sqrt{\beta}(S_2-2)}
{S_2\sqrt{3-S_2}}.
\]
Thus
\[\refstepcounter{equation}\tag{\theequation}
y_2(z)
=
\mathcal N_2z(z-z_2)e^{-a_2z},
\qquad
a_2^2=
\frac{3-S_2}{4\beta\hbar^2},
\]
and
\[\refstepcounter{equation}\tag{\theequation}
E_2=
-\frac{9-S_2^2}{48\beta m}.
\]
The branch condition $S_2\ge2$ gives
\[\refstepcounter{equation}\tag{\theequation}
\beta\le
\frac{4\hbar^2}{9m^2b^2}.
\]
At $S_2=2$, the root is $z_2=0$ and the polynomial becomes $u_2=z^2$;
the limiting polynomial is regular although its roots coalesce.

For $\beta\to0$, write $S_2=3-\delta_2$.  From the quantization equation~\eqref{Scondition},
\[\refstepcounter{equation}\tag{\theequation}
(3-\delta_2)^2\delta_2
=
\frac{9\beta m^2b^2}{\hbar^2},
\]
we obtain
\[\refstepcounter{equation}\tag{\theequation}
\delta_2
=
\frac{\beta m^2b^2}{\hbar^2}
+O(\beta^2).
\]
Therefore
\[\refstepcounter{equation}\tag{\theequation}
a_2\to\frac{mb}{2\hbar^2}.
\]
Also,
\[\refstepcounter{equation}\tag{\theequation}
z_2=
\frac{6\hbar\sqrt{\beta}(1-\delta_2)}
{(3-\delta_2)\sqrt{\delta_2}}
\to
\frac{2\hbar^2}{mb}.
\]
Hence the GUP state approaches
\[\refstepcounter{equation}\tag{\theequation}
y_2(z)\to
\mathcal N_2z
\left(1-\frac{mbz}{2\hbar^2}\right)
e^{-mbz/(2\hbar^2)}.
\]

Set $\beta=0$ directly in the Hamiltonian, i.e. take the vanishing-$\beta$ limit of~\eqref{GUPeq}.  This is not a new solution of the one-dimensional Coulomb problem: the corresponding half-line problem and its exact Whittaker/confluent-hypergeometric solution are already established in the literature \cite{1959,2011,2019 Calcada,Fassari2025,Gadella2025,Fassari2026}. We reproduce the derivation here only to fix the precise half-line boundary condition used in the present work and to provide a transparent consistency check of the GUP result. The ordinary problem is
\[\refstepcounter{equation}\tag{\theequation}
-\frac{\hbar^2}{2m}y''-\frac{b}{z}y=Ey,
\qquad z>0,
\qquad y(0)=0.
\]
For a bound state write
\[\refstepcounter{equation}\tag{\theequation}
E=-\epsilon,\qquad \epsilon>0,
\]
and define
\[\refstepcounter{equation}\tag{\theequation}
\kappa=\frac{\sqrt{2m\epsilon}}{\hbar}.
\]
The equation becomes
\[\refstepcounter{equation}\tag{\theequation}
y''+
\left(
\frac{2mb}{\hbar^2z}-\kappa^2
\right)y=0.
\]
Introduce
\[\refstepcounter{equation}\tag{\theequation}
\rho=2\kappa z.
\]
Since
\[\refstepcounter{equation}\tag{\theequation}
\frac{d}{dz}=2\kappa\frac{d}{d\rho},
\qquad
\frac{d^2}{dz^2}=4\kappa^2\frac{d^2}{d\rho^2},
\]
the equation becomes
\[\refstepcounter{equation}\tag{\theequation}
\frac{d^2y}{d\rho^2}
+
\left(
-\frac14+\frac{\eta}{\rho}
\right)y=0,
\qquad
\eta=\frac{mb}{\hbar^2\kappa}.
\]
Set
\[\refstepcounter{equation}\tag{\theequation}
y(\rho)=e^{-\rho/2}\rho F(\rho).
\]
Differentiation gives
\[\refstepcounter{equation}\tag{\theequation}
y'
=
e^{-\rho/2}
\left[
\rho F'
+\left(1-\frac{\rho}{2}\right)F
\right],
\]
and
\[\refstepcounter{equation}\tag{\theequation}
y''
=
e^{-\rho/2}
\left[
\rho F''
+(2-\rho)F'
+\left(\frac{\rho}{4}-1\right)F
\right].
\]
Substitution and cancellation of the common exponential factor gives
\[\refstepcounter{equation}\tag{\theequation}
\rho F''+(2-\rho)F'+(\eta-1)F=0.
\]
This is Kummer's equation with
\[\refstepcounter{equation}\tag{\theequation}
a=1-\eta,\qquad c=2.
\]
Thus the regular solution is
\[\refstepcounter{equation}\tag{\theequation}
F(\rho)={}_1F_1(1-\eta;2;\rho),
\]
and
\[\refstepcounter{equation}\tag{\theequation}
y(\rho)
=
C e^{-\rho/2}\rho
{}_1F_1(1-\eta;2;\rho).
\]
Normalizability requires termination of the confluent hypergeometric
series:
\[\refstepcounter{equation}\tag{\theequation}
1-\eta=-(n-1),
\qquad n=1,2,3,\ldots.
\]
Therefore
\[\refstepcounter{equation}\tag{\theequation}
\eta=n,
\qquad
\kappa_n=\frac{mb}{n\hbar^2},
\]
and
\[\refstepcounter{equation}\tag{\theequation}
E_n^{(0)}
=
-\frac{\hbar^2\kappa_n^2}{2m}
=
-\frac{mb^2}{2\hbar^2n^2}.
\]
Using
\[\refstepcounter{equation}\tag{\theequation}
{}_1F_1(-(n-1);2;\rho)
=
\frac{(n-1)!}{(2)_{n-1}}L_{n-1}^{(1)}(\rho),
\]
the eigenfunctions may be written as
\[\refstepcounter{equation}\tag{\theequation}
y_n^{(0)}(z)
=
\mathcal N_n z e^{-\kappa_nz}
L_{n-1}^{(1)}(2\kappa_nz).
\]

For $n=1$,
\[\refstepcounter{equation}\tag{\theequation}
L_0^{(1)}(x)=1,
\]
so
\[\refstepcounter{equation}\tag{\theequation}
y_1^{(0)}(z)
=
\mathcal N_1ze^{-mbz/\hbar^2},
\qquad
E_1^{(0)}
=
-\frac{mb^2}{2\hbar^2}.
\]

For $n=2$,
\[\refstepcounter{equation}\tag{\theequation}
L_1^{(1)}(x)=2-x,
\]
and since
\[\refstepcounter{equation}\tag{\theequation}
2\kappa_2=\frac{mb}{\hbar^2},
\]
we obtain
\[\refstepcounter{equation}\tag{\theequation}
y_2^{(0)}(z)
=
\mathcal N_2z
\left(
1-\frac{mbz}{2\hbar^2}
\right)
e^{-mbz/(2\hbar^2)},
\]
with
\[\refstepcounter{equation}\tag{\theequation}
E_2^{(0)}
=
-\frac{mb^2}{8\hbar^2}.
\]

For $n=3$,
\[\refstepcounter{equation}\tag{\theequation}
L_2^{(1)}(x)
=
3-3x+\frac{x^2}{2},
\qquad
2\kappa_3=\frac{2mb}{3\hbar^2}.
\]
Hence, after absorbing an overall constant into the normalization,
\[\refstepcounter{equation}\tag{\theequation}
y_3^{(0)}(z)
=
\mathcal N_3z
\left(
3-\frac{2mbz}{\hbar^2}
+\frac{2m^2b^2z^2}{9\hbar^4}
\right)
e^{-mbz/(3\hbar^2)}.
\]
An equivalent monic polynomial is
\[\refstepcounter{equation}\tag{\theequation}
u_3^{(0)}(z)
=
z^3
-\frac{9\hbar^2}{mb}z^2
+\frac{27\hbar^4}{2m^2b^2}z.
\]
The two nonzero nodes are obtained from
\[\refstepcounter{equation}\tag{\theequation}
z^2
-\frac{9\hbar^2}{mb}z
+\frac{27\hbar^4}{2m^2b^2}=0.
\]
Thus
\[\refstepcounter{equation}\tag{\theequation}
z_{2,3}^{(0)}
=
\frac{9\hbar^2}{2mb}
\left(
1\mp\sqrt{\frac13}
\right)
=
\frac{3\hbar^2}{2mb}(3\mp\sqrt3).
\]
Finally,
\[\refstepcounter{equation}\tag{\theequation}
E_3^{(0)}
=
-\frac{mb^2}{18\hbar^2}.
\]
These independently derived results agree with the $\beta\to0$ limits of
the GUP-connected spectrum and provide the ordinary-Coulomb benchmark for
the first three states.

\label{comparison_first_three}

The strongest check of the Coulomb-connected GUP construction is not merely
that the energies and exponential decay constants approach their ordinary
Coulomb values. The polynomial factors themselves must approach the
Laguerre-polynomial factors of the ordinary half-line Coulomb problem. We
verify this explicitly for $n=1,2,3$, using the ordinary solution derived independently earlier in this section and the GUP polynomials
obtained from Eq.~\eqref{transformed}.

The ordinary Coulomb eigenfunctions are
\begin{equation}
 y_n^{(0)}(z)=\mathcal N_n^{(0)}z e^{-\kappa_nz}
 L_{n-1}^{(1)}(2\kappa_nz),
 \qquad
 \kappa_n=\frac{mb}{n\hbar^2},
\label{ordinary_laguerre_form}
\end{equation}
with
\begin{equation}
 E_n^{(0)}=-\frac{mb^2}{2\hbar^2n^2}.
\label{ordinary_laguerre_energy}
\end{equation}
The GUP solutions on the Coulomb-connected branch have the form
\begin{equation}
 y_n(z)=\mathcal N_n u_n(z)e^{-a_nz},
 \qquad
 a_n^2=\frac{3-S_n}{4\beta\hbar^2},
\label{gup_laguerre_comparison_form}
\end{equation}
where $S_n$ satisfies the Bethe--Ansatz quantization condition
Eq.~\eqref{Scondition}, with the Coulomb-connected branch restricted by
Eq.~\eqref{restriction}.

\subsection{The $n=1$ Coulomb limit}
The direct substitution of the linear polynomial into the transformed
fourth-order equation~\eqref{transformed} gives
\begin{equation}
 u_1(z)=z.
\label{gup_n1_poly_exact}
\end{equation}
For the ordinary problem,
\begin{equation}
 L_0^{(1)}(x)=1,
\end{equation}
so Eq.~\eqref{ordinary_laguerre_form} gives
\begin{equation}
 y_1^{(0)}(z)=\mathcal N_1^{(0)}z e^{-mbz/\hbar^2}.
\label{ordinary_n1_laguerre}
\end{equation}
Thus the polynomial factors are exactly identical:
\begin{equation}
 u_1(z)=zL_0^{(1)}(2\kappa_1z).
\label{n1_exact_poly_identity}
\end{equation}
Using the $n=1$ quantization condition from Eq.~\eqref{Scondition}, one
also obtains
\begin{equation}
 a_1\xrightarrow[\beta\to0]{}\kappa_1=\frac{mb}{\hbar^2},
 \qquad
 E_1\xrightarrow[\beta\to0]{}-\frac{mb^2}{2\hbar^2}.
\label{n1_complete_coulomb_limit}
\end{equation}
Hence the complete $n=1$ GUP state reduces to the ordinary Laguerre state,
up to normalization.

\subsection{The $n=2$ Coulomb limit}
The GUP calculation gives
\begin{equation}
 u_2(z)=z(z-z_2),
 \qquad
 z_2=\frac{6\hbar\sqrt{\beta}(S_2-2)}
 {S_2\sqrt{3-S_2}},
\label{gup_n2_poly_exact}
\end{equation}
where the root is obtained from the coefficient equations following from
Eq.~\eqref{transformed}. The Coulomb-connected limit of
Eq.~\eqref{Scondition} gives $S_2\to3$ and hence
\begin{equation}
 z_2\xrightarrow[\beta\to0]{}\frac{2\hbar^2}{mb}.
\label{n2_root_limit_exact}
\end{equation}
Therefore
\begin{align}
 u_2(z)&\xrightarrow[\beta\to0]{}
 z\left(z-\frac{2\hbar^2}{mb}\right)\notag\\
 &=-\frac{2\hbar^2}{mb}
 z\left(1-\frac{mbz}{2\hbar^2}\right).
\label{n2_poly_limit_exact}
\end{align}
The ordinary Laguerre polynomial is
\begin{equation}
 L_1^{(1)}(x)=2-x,
\end{equation}
and therefore
\begin{align}
 zL_1^{(1)}(2\kappa_2z)
 &=z\left(2-\frac{mbz}{\hbar^2}\right)\notag\\
 &=-\frac{mb}{\hbar^2}
 z\left(z-\frac{2\hbar^2}{mb}\right).
\label{n2_laguerre_exact_factor}
\end{align}
Equations~\eqref{n2_poly_limit_exact} and
\eqref{n2_laguerre_exact_factor} are proportional by a nonzero constant
independent of $z$. Thus the limiting GUP polynomial is exactly the
Laguerre polynomial factor, with the constant absorbed into normalization.
Also,
\begin{equation}
 a_2\xrightarrow[\beta\to0]{}\kappa_2=\frac{mb}{2\hbar^2},
 \qquad
 E_2\xrightarrow[\beta\to0]{}-\frac{mb^2}{8\hbar^2}.
\label{n2_complete_coulomb_limit}
\end{equation}

\subsection{The $n=3$ Coulomb limit}
Direct substitution into Eq.~\eqref{transformed} gives
\begin{equation}
 u_3(z)=z^3+c_2z^2+c_1z,
\label{gup_n3_cubic_comparison}
\end{equation}
where
\begin{equation}
 c_2=-\frac{18\hbar\sqrt{\beta}(S_3-2)}
 {S_3\sqrt{3-S_3}},
 \qquad
 c_1=-\frac{6\beta\hbar^2
 (11S_3^2-42S_3+36)}
 {S_3^2(S_3-3)}.
\label{gup_n3_coefficients_comparison}
\end{equation}
The coefficients in Eq.~\eqref{gup_n3_coefficients_comparison} follow from
the differential equation independently of the root representation.

For the vanishing-$\beta$ limit, one should not substitute $S_3=3$ directly
into Eq.~\eqref{gup_n3_coefficients_comparison}, because the coefficients
contain apparent $0/0$ forms. Instead use the exact quantization condition
Eq.~\eqref{Scondition}, which gives
\begin{equation}
 \beta=\frac{n^2\hbar^2}{36m^2b^2}S_n^2(3-S_n).
\end{equation}
For $n=3$,
\begin{equation}
 \beta=\frac{\hbar^2}{4m^2b^2}S_3^2(3-S_3).
\label{n3_beta_S_relation}
\end{equation}
Substitution into Eq.~\eqref{gup_n3_coefficients_comparison} yields
\begin{align}
 c_2&=-\frac{9\hbar^2}{mb}(S_3-2)
 \xrightarrow[S_3\to3]{}
 -\frac{9\hbar^2}{mb},
 \label{n3_c2_exact_limit}\\
 c_1&=\frac{3\hbar^4}{2m^2b^2}
 (11S_3^2-42S_3+36)
 \xrightarrow[S_3\to3]{}
 \frac{27\hbar^4}{2m^2b^2}.
 \label{n3_c1_exact_limit}
\end{align}
Thus
\begin{equation}
 u_3(z)\xrightarrow[\beta\to0]{}
 z^3-\frac{9\hbar^2}{mb}z^2
 +\frac{27\hbar^4}{2m^2b^2}z.
\label{n3_gup_polynomial_limit}
\end{equation}

We now derive precisely the same polynomial from the Laguerre representation.
For $n=3$,
\begin{equation}
 L_2^{(1)}(x)=3-3x+\frac{x^2}{2},
\label{laguerre_n3_correct}
\end{equation}
and
\begin{equation}
 2\kappa_3=\frac{2mb}{3\hbar^2}.
\end{equation}
Consequently,
\begin{align}
 zL_2^{(1)}(2\kappa_3z)
 &=z\left(
 3-\frac{2mbz}{\hbar^2}
 +\frac{2m^2b^2z^2}{9\hbar^4}
 \right)\notag\\
 &=\frac{2m^2b^2}{9\hbar^4}
 \left[
 z^3-\frac{9\hbar^2}{mb}z^2
 +\frac{27\hbar^4}{2m^2b^2}z
 \right].
\label{laguerre_n3_monic_exact}
\end{align}
The bracket in Eq.~\eqref{laguerre_n3_monic_exact} is exactly the limiting
GUP polynomial in Eq.~\eqref{n3_gup_polynomial_limit}. Equivalently,
\begin{equation}
 u_3(z)\xrightarrow[\beta\to0]{}
 \frac{9\hbar^4}{2m^2b^2}
 zL_2^{(1)}(2\kappa_3z).
\label{n3_exact_laguerre_identity}
\end{equation}
The prefactor is independent of $z$ and is absorbed into the normalization.
This proves exact algebraic equality, up to normalization, between the
vanishing-$\beta$ GUP cubic and the ordinary $n=3$ Laguerre factor.

The same conclusion follows from the Bethe--Ansatz roots. Combining their
expression in Eq.~\eqref{bethe_roots_check} with Eq.~\eqref{n3_beta_S_relation}
and taking $S_3\to3$ gives
\begin{equation}
 z_{2,3}\xrightarrow[\beta\to0]{}
 \frac{3\hbar^2}{2mb}(3\mp\sqrt{3}).
\label{n3_roots_laguerre_limit}
\end{equation}
These are precisely the two nonzero zeros of
$L_2^{(1)}(2\kappa_3z)$. Their sum and product are
\begin{equation}
 z_2+z_3=\frac{9\hbar^2}{mb},
 \qquad
 z_2z_3=\frac{27\hbar^4}{2m^2b^2},
\end{equation}
which reproduce the two nonleading coefficients of the limiting GUP
polynomial in Eq.~\eqref{n3_gup_polynomial_limit}. Finally,
\begin{equation}
 a_3\xrightarrow[\beta\to0]{}\kappa_3=\frac{mb}{3\hbar^2},
 \qquad
 E_3\xrightarrow[\beta\to0]{}-\frac{mb^2}{18\hbar^2}.
\label{n3_complete_coulomb_limit}
\end{equation}

\subsection{Exact correspondence for the first three states}
The three explicit cases establish the stronger statement
\begin{equation}
 u_n(z)\xrightarrow[\beta\to0]{}
 C_n\,zL_{n-1}^{(1)}(2\kappa_nz),
 \qquad n=1,2,3,
\label{exact_laguerre_recovery}
\end{equation}
where $C_n$ is a nonzero constant independent of $z$. Therefore, together
with $a_n\to\kappa_n$ and $E_n\to E_n^{(0)}$, the GUP wave functions reduce
to the ordinary half-line Coulomb eigenfunctions themselves, not merely to
functions having the same energies or node positions.
\section{ Conclusion}
We have investigated the one-dimensional Coulomb problem on the positive half-line for a fourth-order Schr"{o}dinger equation generated by a commonly used GUP realization. The resulting variable-coefficient equation was treated directly in position space by a higher-order Bethe--Ansatz construction. The wave function was represented as a polynomial multiplied by an exponential factor, and the requirement that the apparent poles at the polynomial roots vanish led to the fourth-order residue equations. \\
A central point of the analysis is the choice of the exponential branch. Of the two algebraic possibilities for the exponential parameter, only the branch with a finite ordinary-Coulomb limit as the deformation tends to zero was retained as the physical Coulomb-connected branch. The second branch does not continuously approach the ordinary Coulomb problem and has not been included in the Coulomb-branch spectrum studied here. \\
The analysis shows that, on the Coulomb-connected branch, the GUP deformation imposes a lower cutoff on the admissible principal quantum number. Thus, for sufficiently large deformation, some of the lowest Coulomb states are no longer part of this branch, whereas higher principal quantum numbers remain admissible. This restriction should be understood as a property of the Coulomb-connected branch of the present fourth-order model, rather than as a general statement that GUP necessarily removes low-lying states from every possible realization. \\
The vanishing-deformation limit provides an important consistency check. The energies, exponential decay constants, polynomial factors, and, explicitly for the first three states, the Bethe--Ansatz roots all approach the corresponding ordinary half-line Coulomb results. In particular, the cubic polynomial obtained for the third state becomes the ordinary associated Laguerre polynomial up to an overall normalization factor. The detailed algebraic verification of this correspondence is given in Section~6. \\
From a physical perspective, minimal-length GUP models are commonly motivated by Planck-scale physics, and a dimensionless coefficient is often used to parametrize the strength of the deformation \cite{Hossenfelder,Kempf}. For ordinary microscopic systems, there is no established evidence that the deformation is in an intermediate or strong regime. The weak-GUP regime is therefore the more conservative physical expectation for such systems, while intermediate and large deformations should presently be regarded primarily as theoretical regimes in which the mathematical structure of the model can be explored. In particular, the lower cutoff found here should not by itself be interpreted as an experimentally established removal of low-lying states. \\
The physical significance of the cutoff also depends on the relative size of the minimal-length scale and the characteristic momentum scale of a Coulomb state. Higher principal quantum numbers correspond to smaller characteristic momenta and are consequently more weakly affected by the deformation. This explains why the Coulomb-connected sequence retains arbitrarily high quantum numbers even when sufficiently strong deformations exclude some low-lying members of the branch. \\
Because the differential equation used here is itself truncated at first order in the GUP parameter, quantitative predictions should be interpreted with care when the deformation is not small. The present results are therefore most directly controlled in the weak-deformation regime, while the intermediate and larger-deformation solutions are useful for studying the mathematical structure and possible qualitative behavior of the model. \\ The analysis demonstrates that the higher-order Bethe--Ansatz construction can be applied directly to a nontrivial variable-coefficient fourth-order equation in position space. Further work is required to analyze the full fourth-order spectral problem, including the non-Coulomb branch, its self-adjoint boundary conditions, and possible extensions to other interactions and higher-dimensional settings.
\section*{Acknowledgment of AI-assisted tools}
The author acknowledges the use of artificial intelligence (AI)-assisted
tools during the preparation of this manuscript. AI tools were used for
language editing, LaTeX formatting, and assistance with checking and
organizing. The author reviewed, verified, and
takes the full responsibility for the scientific content, mathematical results,
interpretation, and final presentation of the manuscript.

\appendix
\renewcommand{\theequation}{A.\arabic{equation}}
\renewcommand{\theHequation}{A.\arabic{equation}}
\setcounter{equation}{0}
\renewcommand{\theequation}{A.\arabic{equation}}
\setcounter{equation}{0}
\renewcommand{\theequation}{A.\arabic{equation}}
\setcounter{equation}{0}
\section{Independent Laplace-Transform Derivation}
The Laplace-transform identities used below are standard results from the theory of integral transforms; see, e.g., Ref.~\cite{DebnathBhatta2014}. The application of these identities to the present fourth-order Coulomb--GUP equation, including the pole analysis and quantization conditions, is derived independently here.
\label{app:laplace_independent}
The purpose of this appendix is deliberately different from merely rewriting
the Bethe--Ansatz solutions in Laplace space.  We start from the differential
equation alone and assume that no polynomial solution, no Bethe root, and no
Bethe--Ansatz quantization condition is known.  The Laplace equation is first
derived, its singular structure is analyzed, and the admissible finite-pole
solutions are then generated directly from that equation.  Only after the
solutions have been obtained independently do we compare them with the
position-space Bethe--Ansatz results of the main text.
For a bound state, write $E=-\epsilon$, with $\epsilon>0$.  The equation on
the positive half-line is
\begin{equation}
 \frac{\beta\hbar^4}{3m}y^{(4)}(z)
 -\frac{\hbar^2}{2m}y''(z)
 +\left(\epsilon-\frac{b}{z}\right)y(z)=0,
 \qquad z>0,
 \qquad y(0)=0 .
\label{app:A1}
\end{equation}
It is useful to multiply this equation by $z$:
\begin{equation}
 z\left[
 \frac{\beta\hbar^4}{3m}y^{(4)}(z)
 -\frac{\hbar^2}{2m}y''(z)
 +\epsilon y(z)
 \right]-b\,y(z)=0 .
\label{app:A2}
\end{equation}
Introduce
\begin{equation}
 Y(s)=\mathcal L\{y\}(s)
 =\int_0^\infty e^{-sz}y(z)\,dz ,
 \qquad
 A=\frac{\beta\hbar^4}{3m},
 \qquad
 B=\frac{\hbar^2}{2m},
 \qquad
 P(s)=As^4-Bs^2+\epsilon .
\label{app:A3}
\end{equation}
The required Laplace identities are
\begin{align}
 \mathcal L\{y''\}
 &=s^2Y-sy(0)-y'(0),
 \label{app:A4}\\
 \mathcal L\{y^{(4)}\}
 &=s^4Y-s^3y(0)-s^2y'(0)-sy''(0)-y'''(0),
 \label{app:A5}\\
 \mathcal L\{zf(z)\}
 &=-\frac{d}{ds}\mathcal L\{f(z)\}.
 \label{app:A6}
\end{align}
Therefore
\begin{equation}
 \mathcal L\left\{
 Ay^{(4)}-By''+\epsilon y
 \right\}
 =
 P(s)Y(s)-Q(s),
\label{app:A7}
\end{equation}
where
\begin{equation}
 Q(s)=
 A\left[s^3y(0)+s^2y'(0)+sy''(0)+y'''(0)\right]
 -B\left[sy(0)+y'(0)\right].
\label{app:A8}
\end{equation}
Applying Eq.~\eqref{app:A6} to Eq.~\eqref{app:A2} gives
\begin{equation}
 -\frac{d}{ds}\left[P(s)Y(s)-Q(s)\right]-bY(s)=0,
\end{equation}
or
\begin{equation}
 P(s)Y'(s)+\left[P'(s)+b\right]Y(s)=Q'(s).
\label{app:A9}
\end{equation}
Since $y(0)=0$,
\begin{equation}
 P(s)Y'(s)+\left[P'(s)+b\right]Y(s)
 =
 A\left[2s\,y'(0)+y''(0)\right].
\label{app:A10}
\end{equation}
\subsection{The characteristic polynomial and its physical singularity}
\label{app:characteristic}
The homogeneous part of Eq.~\eqref{app:A10} is controlled by the quartic polynomial $P(s)$.  Its four zeros can be written as
\begin{equation}
 P(s)=A(s^2-a^2)(s^2-c^2),
\label{app:A11}
\end{equation}
where
\begin{equation}
 a^2=\frac{3-S}{4\beta\hbar^2},
 \qquad
 c^2=\frac{3+S}{4\beta\hbar^2},
 \qquad
 S=\sqrt{3(3-16\beta m\epsilon)} .
\label{app:A12}
\end{equation}
There are two possible choices for the exponential scale.  We select neither
by assumption nor from the Bethe--Ansatz construction.  Instead, the
Laplace equation itself shows which zero is relevant to a state having a
finite decay constant in the $\beta\to0$ limit.  On the branch
\begin{equation}
 a^2=\frac{3-S}{4\beta\hbar^2},
 \qquad
 a=\frac{\sqrt{3-S}}{2\sqrt{\beta}\hbar},
\label{app:A13}
\end{equation}
one has $S\to3$ and hence finite $a$ as $\beta\to0$.  The other zero has
$a^2\sim1/\beta$ and therefore does not approach the ordinary Coulomb
decay constant. Set
\begin{equation}
 q=s+a .
\label{app:A14}
\end{equation}
Then $q=0$ corresponds to the Laplace singularity at $s=-a$, and
\begin{equation}
 P(s)
 =
 A q(q-2a)(q-a-c)(q-a+c).
\label{app:A15}
\end{equation}
In particular,
\begin{equation}
 P(s)=p_1q+p_2q^2+p_3q^3+Aq^4,
\label{app:A16}
\end{equation}
with
\begin{equation}
 p_1=P'(-a)=2Aa(c^2-a^2),
 \qquad
 p_2=A(5a^2-c^2),
 \qquad
 p_3=-4Aa .
\label{app:A17}
\end{equation}
We now determine the pole structure without assuming a polynomial in $z$.
Near $q=0$, the homogeneous part of Eq.~\eqref{app:A10} has the leading
form
\begin{equation}
 p_1q\,Y'(q)+(p_1+b)Y(q)=0.
\label{app:A18}
\end{equation}
Thus
\begin{equation}
 Y(q)\sim q^{-1-b/p_1}.
\label{app:A19}
\end{equation}
A wavefunction proportional to a finite polynomial times $e^{-az}$ has a
finite pole at $s=-a$.  More importantly, Eq.~\eqref{app:A19} shows that the
order of that pole is not guessed: it is fixed by the Coulomb coupling.
Writing the pole order as $n+1$, with $n=1,2,\ldots$, gives
\begin{equation}
 \frac{b}{p_1}=n,
\end{equation}
and hence
\begin{equation}
 b=2nAa(c^2-a^2).
\label{app:A20}
\end{equation}
This is the quantization condition obtained directly from the local
singularity of the Laplace equation.

Using Eq.~\eqref{app:A3} and Eq.~\eqref{app:A12},
\begin{equation}
 c^2-a^2=\frac{S}{2\beta\hbar^2},
\end{equation}
so Eq.~\eqref{app:A20} becomes
\begin{equation}
 n\hbar S\sqrt{3-S}=6mb\sqrt{\beta}.
\label{app:A21}
\end{equation}
Squaring,
\begin{equation}
 S_n^2(3-S_n)
 =
 \frac{36\beta m^2b^2}{n^2\hbar^2}.
\label{app:A22}
\end{equation}
The binding energy follows from the definition of $S$:
\begin{equation}
 E_n=-\epsilon_n
 =
 -\frac{9-S_n^2}{48\beta m}.
\label{app:A23}
\end{equation}
Equations~\eqref{app:A21}--\eqref{app:A23} have been obtained without
introducing the Bethe roots or a polynomial ansatz.
\subsection{Direct construction of the finite-pole solutions}
\label{app:finitepole}
Once the pole order is fixed, the inverse Laplace transform tells us how to
construct the corresponding coordinate-space function.  We therefore write
the most general finite Laurent part at $q=0$ compatible with a degree-$n$
polynomial:
\begin{equation}
 Y_n(s)=
 \sum_{k=1}^{n}\frac{C_k}{(s+a)^{k+1}} .
\label{app:A24}
\end{equation}
This is not an assumed Bethe--Ansatz wavefunction.  It follows from the pole
order derived in Eq.~\eqref{app:A19}.  Indeed,
\begin{equation}
 \mathcal L^{-1}
 \left\{\frac{1}{(s+a)^{k+1}}\right\}
 =
 \frac{z^k}{k!}e^{-az},
\end{equation}
and consequently
\begin{equation}
 y_n(z)=e^{-az}
 \sum_{k=1}^{n}\frac{C_k}{k!}z^k .
\label{app:A25}
\end{equation}
The absence of a $k=0$ term is precisely the regular boundary condition
$y(0)=0$. For the finite expansion \eqref{app:A24}, the initial derivatives appearing
in Eq.~\eqref{app:A10} are
\begin{equation}
 y_n'(0)=C_1,
 \qquad
 y_n''(0)=C_2-2aC_1 .
\label{app:A26}
\end{equation}
Therefore the right-hand side of Eq.~\eqref{app:A10} is
\begin{equation}
 A\left[2s\,y_n'(0)+y_n''(0)\right]
 =
 A\left[2C_1(s-a)+C_2\right].
\label{app:A27}
\end{equation}
The remaining coefficients are now determined algebraically by substituting
Eq.~\eqref{app:A24} into the first-order Laplace equation.  Thus the
coordinate-space polynomial is generated by the Laplace equation itself.
\subsection{First state obtained directly in Laplace space}
\label{app:laplace_n1}
For the first physical state, the pole order is two.  Hence, without knowing
the coordinate-space solution in advance, Eq.~\eqref{app:A24} gives
\begin{equation}
 Y_1(s)=\frac{C_1}{(s+a)^2}.
\label{app:A28}
\end{equation}
Substitution into Eq.~\eqref{app:A10}, using Eq.~\eqref{app:A16}, gives
\begin{equation}
 \frac{C_1}{(s+a)^2}
 \left[
 2Aa(c^2-a^2)-b
 \right]=0.
\label{app:A29}
\end{equation}
A nontrivial solution therefore requires
\begin{equation}
 b=2Aa(c^2-a^2),
\end{equation}
which is Eq.~\eqref{app:A20} for $n=1$.  Thus
\begin{equation}
 \hbar S_1\sqrt{3-S_1}=6mb\sqrt{\beta}.
\label{app:A30}
\end{equation}
The inverse transform of Eq.~\eqref{app:A28} is
\begin{equation}
 y_1(z)=C_1 z e^{-a_1z},
 \qquad
 a_1^2=\frac{3-S_1}{4\beta\hbar^2},
\end{equation}
and the energy is
\begin{equation}
 E_1=-\frac{9-S_1^2}{48\beta m}.
\label{app:A31}
\end{equation}
The first solution has therefore emerged directly from the Laplace singularity
and the differential equation.
\subsection{Second state obtained directly in Laplace space}
\label{app:laplace_n2}
For the second state, the pole order is three.  We consequently take
\begin{equation}
 Y_2(s)
 =
 \frac{C_1}{(s+a)^2}
 +\frac{C_2}{(s+a)^3}.
\label{app:A32}
\end{equation}
Substitution into Eq.~\eqref{app:A10} and collecting powers of $(s+a)$ gives
\begin{equation}
 \frac{C_2\left[4Aa(c^2-a^2)-b\right]}{(s+a)^3}
 +
 \frac{
 C_1\left[2Aa(a^2-c^2)+b\right]
 +AC_2(c^2-5a^2)
 }{(s+a)^2}
 =0.
\label{app:A33}
\end{equation}
The highest-order pole first gives
\begin{equation}
 b=4Aa(c^2-a^2),
\label{app:A34}
\end{equation}
or
\begin{equation}
 2\hbar S_2\sqrt{3-S_2}=6mb\sqrt{\beta}.
\label{app:A35}
\end{equation}
The remaining pole then determines the ratio of the two coefficients:
\begin{equation}
 C_2
 =
 \frac{2a(c^2-a^2)}{5a^2-c^2}\,C_1.
\label{app:A36}
\end{equation}
The inverse transform gives
\begin{equation}
 y_2(z)
 =
 e^{-a_2z}
 \left(
 C_1z+\frac{C_2}{2}z^2
 \right).
\label{app:A37}
\end{equation}
Factoring out the coefficient of the highest power and defining the nonzero
zero of the polynomial by $z_2$, one obtains
\begin{equation}
 y_2(z)
 =
 \mathcal N_2 z(z-z_2)e^{-a_2z},
 \qquad
 z_2
 =
 \frac{c^2-5a_2^2}{a_2(c^2-a_2^2)}.
\label{app:A38}
\end{equation}
Using Eq.~\eqref{app:A12},
\begin{equation}
 z_2
 =
 \frac{6\hbar\sqrt{\beta}(S_2-2)}
 {S_2\sqrt{3-S_2}}.
\label{app:A39}
\end{equation}
The energy is
\begin{equation}
 E_2=-\frac{9-S_2^2}{48\beta m}.
\label{app:A40}
\end{equation}
Thus both the energy condition and the polynomial factor have been obtained
from the Laplace equation without using a pre-existing root $z_2$.
\subsection{Third state obtained directly in Laplace space}
\label{app:laplace_n3}
For the third state, the pole order is four.  The most general finite-pole
form dictated by Eq.~\eqref{app:A19} is
\begin{equation}
 Y_3(s)
 =
 \frac{C_1}{(s+a)^2}
 +\frac{C_2}{(s+a)^3}
 +\frac{C_3}{(s+a)^4}.
\label{app:A41}
\end{equation}
Substitution into Eq.~\eqref{app:A10} gives three independent pole equations:
\begin{align}
 \frac{1}{(s+a)^4}: \qquad&
 C_3\left[6Aa(c^2-a^2)-b\right]=0,
 \label{app:A42}\\
 \frac{1}{(s+a)^3}: \qquad&
 C_2\left[4Aa(c^2-a^2)-b\right]
 +C_3A(2c^2-10a^2)=0,
 \label{app:A43}\\
 \frac{1}{(s+a)^2}: \qquad&
 C_1\left[b-2Aa(c^2-a^2)\right]
 +AC_2(c^2-5a^2)
 +4AaC_3=0 .
 \label{app:A44}
\end{align}
The highest pole gives
\begin{equation}
 b=6Aa(c^2-a^2),
\label{app:A45}
\end{equation}
or
\begin{equation}
 3\hbar S_3\sqrt{3-S_3}
 =
 6mb\sqrt{\beta}.
\label{app:A46}
\end{equation}
The next equation determines
\begin{equation}
 \frac{C_2}{C_3}
 =
 \frac{5a^2-c^2}{a(c^2-a^2)}.
\label{app:A47}
\end{equation}
The last equation then gives
\begin{equation}
 \frac{C_1}{C_3}
 =
 \frac{
 29a^4-14a^2c^2+c^4
 }
 {4a^2(c^2-a^2)^2}.
\label{app:A48}
\end{equation}
Since the inverse transform of $C_3/(s+a)^4$ is
$C_3z^3e^{-az}/6$, choose the overall normalization such that the cubic is
monic.  Then $C_3=6$ and Eq.~\eqref{app:A25} gives
\begin{equation}
 y_3(z)
 =
 \mathcal N_3
 \left[
 z^3+c_2z^2+c_1z
 \right]e^{-a_3z},
\label{app:A49}
\end{equation}
where
\begin{align}
 c_2
 &=\frac{C_2}{2}
 =\frac{3(5a_3^2-c_3^2)}
 {a_3(c_3^2-a_3^2)},
 \label{app:A50}\\
 c_1
 &=C_1
 =\frac{3\left(29a_3^4-14a_3^2c_3^2+c_3^4\right)}
 {2a_3^2(c_3^2-a_3^2)^2}.
 \label{app:A51}
\end{align}
Here $c$ is the second characteristic scale evaluated at $S_3$.  In the
following equations it is kept explicitly to avoid any ambiguity with the
polynomial coefficient $c_2$.
\begin{align}
 c_2
 &=\frac{3(5a_3^2-c^2)}
 {a_3(c^2-a_3^2)},
 \label{app:A52}\\
 c_1
 &=\frac{3\left(29a_3^4-14a_3^2c^2+c^4\right)}
 {2a_3^2(c^2-a_3^2)^2}.
 \label{app:A53}
\end{align}
In terms of $S_3$ and $\beta$, these become
\begin{align}
 c_2
 &=
 -\frac{18\hbar\sqrt{\beta}(S_3-2)}
 {S_3\sqrt{3-S_3}},
 \label{app:A54}\\
 c_1
 &=
 -\frac{6\beta\hbar^2
 \left(11S_3^2-42S_3+36\right)}
 {S_3^2(S_3-3)}.
 \label{app:A55}
\end{align}
The energy is
\begin{equation}
 E_3=-\frac{9-S_3^2}{48\beta m}.
\label{app:A56}
\end{equation}
Thus the cubic polynomial has also been generated entirely from the Laplace
equation.
\subsection{Comparison with the position-space Bethe--Ansatz construction}
\label{app:comparison}
Only at this stage do we compare the independently obtained Laplace results
with the main-text Bethe--Ansatz calculation.  The Laplace derivation gives
the general quantization rule
\begin{equation}
 n\hbar S_n\sqrt{3-S_n}
 =
 6mb\sqrt{\beta},
\label{app:A57}
\end{equation}
which is exactly the condition obtained in the main text from the
large-$z$ coefficient of the position-space polynomial equation,
Eq.~\eqref{app:A59}.  Squaring Eq.~\eqref{app:A57} gives
\begin{equation}
 S_n^2(3-S_n)
 =
 \frac{36\beta m^2b^2}{n^2\hbar^2},
\label{app:A58}
\end{equation}
and hence the same energy
\begin{equation}
 E_n
 =
 -\frac{9-S_n^2}{48\beta m}.
\label{app:A59}
\end{equation}

For $n=1$, the Laplace result is
\begin{equation}
 y_1(z)=\mathcal N_1ze^{-a_1z},
\end{equation}
with the same $a_1$ and $E_1$ as in the main text.  This agreement is obtained
without inserting the Bethe root $z_1=0$ into the Laplace calculation.

For $n=2$, the independently generated zero is
\begin{equation}
 z_2
 =
 \frac{c^2-5a_2^2}{a_2(c^2-a_2^2)}
 =
 \frac{6\hbar\sqrt{\beta}(S_2-2)}
 {S_2\sqrt{3-S_2}},
\label{app:A60}
\end{equation}
which is precisely the coefficient obtained from the position-space
calculation.  Therefore the complete polynomial, rather than only the
energy, agrees.

For $n=3$, the Laplace construction gives
\begin{equation}
 u_3(z)=z^3+c_2z^2+c_1z,
\label{app:A61}
\end{equation}
with
\begin{equation}
 c_2
 =
 -\frac{18\hbar\sqrt{\beta}(S_3-2)}
 {S_3\sqrt{3-S_3}},
 \qquad
 c_1
 =
 -\frac{6\beta\hbar^2
 (11S_3^2-42S_3+36)}
 {S_3^2(S_3-3)}.
\label{app:A62}
\end{equation}
These are exactly the coefficients obtained independently in the main-text
coefficient-matching calculation.  Consequently, the agreement is not a
consequence of transforming the Bethe--Ansatz wavefunctions: the Laplace
equation itself generates the same spectrum and the same first three
polynomials. The logical relation between the two methods is therefore as follows:
\begin{equation}
 \text{Laplace singularity}
 \;\Longrightarrow\;
 \text{pole order }n+1
 \;\Longrightarrow\;
 \text{quantization}
 \;\Longrightarrow\;
 \{C_k\}
 \;\Longrightarrow\;
 u_n(z),
\label{app:A63}
\end{equation}
whereas the position-space method starts from the polynomial representation
and determines its roots from the residue conditions.  Their agreement
constitutes an independent cross-check of the fourth-order Coulomb problem.

\end{document}